\documentclass[twocolumn,showpacs,preprintnumbers,draftclsnofoot]{revtex4}
\usepackage{amsmath}
\usepackage{amssymb}
\usepackage{graphicx}
\usepackage{dcolumn}
\usepackage{bm}
\usepackage{amssymb}
\usepackage{graphicx}
\usepackage{dcolumn}
\usepackage{bm}
\begin{document}

\title{ Chiral Majorana Fermions in two dimensional square lattice antiferromagnet with proximity-induced superconductivity  }
\author{Ma Luo\footnote{Corresponding author:luoma@gpnu.edu.cn} }
\affiliation{School of Optoelectronic Engineering, Guangdong Polytechnic Normal University, Guangzhou 510665, China}

\begin{abstract}

Combination of proximity-induced superconductivity and ferromagnetic exchange field in a two-dimensional square-lattice antiferromagnet with spin-orbit coupling and nonsymmorphic symmetry can induce a topological superconductor phase with chiral Majorana edge states. The lattice model of the Bogoliubov-de Gennes (BdG) Hamiltonian was applied to study the phase diagram of bulks and chiral Majorana edge states in nanoribbons. By numerically studying the phase diagram, we found that the non-uniformity of either the superconducting pairing parameters or the exchange field at the two sublattices is necessary to induce a topological superconductor phase with chiral Majorana edge states. The BdG Chern number of certain topological superconductor phases is $\pm1$ or $\pm3$, such that the corresponding nanoribbons have one or three pairs of chiral Majorana edge states, respectively.

\end{abstract}

\maketitle

\section{Introduction}

Majorana fermions are one of the three types of fundamental fermions in particle physics, which are theoretically predicted by the real solution of the Dirac equation \cite{Majorana1937}. The major feature of the Majorana fermion is that the antiparticle of the Majorana fermion is itself. However, Majorana fermions have not been experimentally observed \cite{Elliott2015}.

On the other hand, in condensed matter physics, Majorana fermion modes have been put forward to mimic the Majorana fermion in particle physics \cite{Elliott2015,Aguado2017}. The antiparticle of an electron above the Fermi level is a hole at the Fermi level. In a superconductor, a quasi-particle is a coherent coupling between an electron and hole. The Majonara zero mode (MZM) is a quasi-particle with equal superposition of electron and hole degrees of freedom and zero energy, such that its antiparticle is itself. Theoretical studies have shown that exchanging two MZM (i.e., braiding) is accompanied by a unitary operation on the wave function, so that the MZM is neither Boson nor Fermion, but non-Abelian anyons. Non-Abelian statistical properties can be applied to robust topological quantum computers \cite{Nayak2008}. As a result, much research has been devoted to finding systems that host MZM \cite{Manisha2017,Zhongbo2018,Yohanes2018,ZhenHuaWang2018,Zhongbo2019,Katharina2020,Yohanes2020,Zhenzhen2020}.

Because the braiding operation is difficult for a localized zero-dimensional MZM, some other proposals suggest implementing the braiding operation based on the chiral Majorana mode at the edge of a two-dimensional heterostructure \cite{Lian2018,Beenakker2019}. A few platforms have been proposed to realize the chiral Majorana mode \cite{DimitrieCulcer20}. Proximity between three dimensional topological insulator and conventional superconductor induce topological superconducting phase at the interface \cite{LiangFu08}. Experimental observation of the two-dimensional topological superconductivity was implemented in the Cr-doped
(Bi,Se)$_{2}$Te$_{3}$ thin films with Nb contacts \cite{experiTITSC1,experiTITSC2}, or the Pb/Co/Si(111) heterostructure \cite{Guissart17}. By inducing superconductivity in two-dimensional systems in the quantum anomalous Hall (QAH) phase, a chiral topological superconductor with Bogoliubov-de Gennes (BdG) Chern number of nonzero could be engineered \cite{XLQi2009,Schnyder2008,XLQi2010}. A few previous studies proposed the graphene/superconductor \cite{Jelena2013,Dutreix2014,Wang2016,ZhenHuaWang2019,Manesco2019,Petra2020} heterostructure as platform for observation of MZM, because of the multiple advantages of the electronic and mechanical properties of graphene \cite{CastroNeto2009}. The magnetic exchange field in graphene can be obtained by doping or by proximity to a ferromagnetic substrate. The additional presence of Rashba spin-orbit coupling (SOC) induces the QAH phase \cite{Zhenhua2010,Zhenhua2014}. However, pristine graphene has weak SOC \cite{Kane2005,Hongki2006,Yugui2007}. Doping \cite{Fufang2007,JunHu2012} or the proximity effect \cite{Zhenhua2010,Emmanuel2009,Jayakumar2014,Abdulrhman2018} can enhance the SOC, but the effect is limited. Doping with magnetic elements can change the atomic structure of graphene, which is not advantageous. In additional, the models require large value of superconducting gap to induce the topological superconducting phase, which is not realistic. As a result, it is reasonable to search for other two-dimensional systems with a large internal SOC so that a topological superconductor phase can be obtained without external doping. One feasible scheme is to combine superconductors and antiferromagnetic topological insulators (AFTI) \cite{Yingyi18,YangPeng19,Beibing22}. Because the three-dimensional materials in the AFTI phase have a large intrinsic SOC, a topological superconducting phase with a large gap could be induced at the interface between the AFTI and regular superconductor. Another scheme for creating a chiral Majorana edge mode is to deposit a lattice of magnetic adatoms on the surface of conventional superconductors with Rashba SOC, as described by the Yu-Shiba-Rusinov lattice model \cite{StephanRachel17,DanielCrawford20,PalacioMorales19}. The coupling between the conventional superconductor and the Zeeman field of the adatom creates Yu-Shiba-Rusinov states, which host chiral Majorana edge modes.

In this study, we propose another scheme that combines a two-dimensional (2D) AFTI and a regular superconductor, which induces a topological superconductor phase in the 2D lattice. The specific materials of 2D AFTI are assumed to be antiferromagnetic (AF) XMnY (X=Sr and Ba, Y=Sn and Pb) quintuple layers (QL) \cite{Chengwang2020,NingMao20}, which have been proposed to be dynamically stable as 2D crystals. The AF XMnY QL is an intrinsic AF with a large SOC and non-symmorphic symmetry \cite{ChengCheng2011,Steve2015}. In a certain phase regime, the system is in the topological insulator phase, with a helical edge state at the edge of the nanoribbon. We studied the heterostructure of AF XMnY QL in proximity to an s-wave superconductor substrate and a ferromagnetic substrate. The presence of a superconductor substrate induces superconducting pairing in AF XMnY QL, which couples the electrons and holes. The additional presence of the ferromagnetic substrate induces a ferromagnetic exchange field \cite{Cardoso2018} in AF XMnY QL, which in turn modifies the phase diagram of the topological superconductor phase. Topological phases with a BdG Chern number of $\pm1$, $\pm2$ or $\pm3$ could be obtained. The lattice model in the particle-hole symmetric Nambu spinor space was applied to study the heterostructure.

The remainder of this paper is organized as follows. II, the theoretical
model of the heterostructure is described. In Sec. III, the phase diagram of AF XMnY QL on a superconductor was studied. In Sec. IV, the effects of an additional ferromagnetic substrate and chiral Majorana edge mode were studied. In Sec. V, and the conclusions are presented.

\section{Theoretical model}

\begin{figure}[tbp]
\scalebox{0.58}{\includegraphics{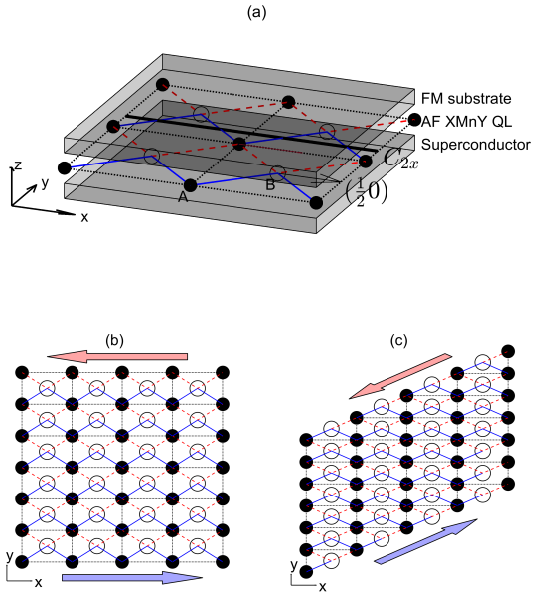}}
\caption{ Schematic of the AF XMnY QL, which is sandwiched between the superconductor substrate and the ferromagnetic substrate. The lattice structure of the AF XMnY QL is indicated by the checkerboard square lattice. The lattice sites of A and B sublattice are indicated by fill and empty dots, respectively. The inter-sublattice hopping between nearest neighboring sites are indicated by red (solid) or blue (dashed) lines, whose strength are $t_{1}$ or $t_{2}$, respectively. The intra-sublattice hopping between nearest neighboring sites are indicated by the black (dotted) lines. The twofold screw symmetry is designated as $C_{2x}$, and a half of the lattice translation along $\hat{x}$ direction is designated as $(\frac{1}{2}0)$. (b) and (c) are schematic of parallel and diagonal nanoribbons, whose periodical direction is along $\hat{x}$ and $\hat{x}+\hat{y}$, respective. The chiral Majorana edge modes are indicated by the thick arrows.  }
\label{figure_system}
\end{figure}

The schematic of the system is shown in Fig. \ref{figure_system}(a). AF XMnY QL is indicated by the checkerboard square lattice in the middle layer. Sublattices A and B are marked by filled and empty dots, respectively. An s-wave superconductor is in proximity to the AF XMnY QL, which induces superconducting pairing in AF XMnY QL. On the other side of AF XMnY QL, a ferromagnetic substrate is in proximity to AF XMnY QL, which induces a ferromagnetic exchange field in AF XMnY QL. The lattice model includes one atomic orbit in each sublattice such that two atomic orbits are included in each unit cell. Including spin-up and spin-down, the Hamiltonian of the four-band tight binding model for bulk is given by \cite{Chengwang2020,ChengCheng2011,Steve2015}
\begin{eqnarray}
&&H_{QL}(\mathbf{k})=[Re(\tilde{M})\tau_{x}-Im(\tilde{M})\tau_{y}]\sigma_{0}\\
&&-2t_{I}(\cos{k_{x}}+\cos{k_{y}})\tau_{z}\sigma_{z}\nonumber\\
&&+t_{R}\tau_{z}(\sigma_{y}\sin{k_{x}}-\sigma_{x}\sin{k_{y}})
+\lambda_{AF}\tau_{z}\sigma_{z}+\lambda_{FM}\tau_{0}\sigma_{z}\nonumber
\end{eqnarray}
where $\tau_{0,x,y,z}$ and $\sigma_{0,x,y,z}$ are the Pauli matrices of the sublattice and spin, respectively, $k_{x,y}$ are the normalized Bloch wave numbers of the lattice, $t_{I}$ is the strength of the intrinsic SOC, $t_{R}$ is the strength of the Rashba SOC, $\lambda_{AF}$ is the strength of the internal antiferromagnetic exchange field of the QL, $\lambda_{FM}$ is the strength of the externally induced ferromagnetic exchange field owing to its proximity to the ferromagnetic substrate \cite{Cardoso2018}. The hopping energies between the nearest neighboring A and B lattice sites, which are marked by the blue (solid) and red (dashed) lines in Fig. \ref{figure_system}(a) is $t_{1}$ and $t_{2}$, so that the parameters of the first term in the Hamiltonian is given as $\tilde{M}=(t_{1}+t_{2}e^{ik_{y}})(1+e^{-ik_{x}})$. To demonstrate the physical model of the AF XMnY QL, the parameters were assumed to be $t_{1}=0.7$ eV, $t_{2}=0.4$ eV, $t_{I}=0.3$ eV, $t_{R}=1$ eV, $\lambda_{AF}=\pm0.4$ eV in the following two sections. The ferromagnetic exchange field is either zero or $\lambda_{FM}=\pm0.2$ eV if a ferromagnetic substrate is absent or present, respectively. The systems with realistic parameters are discussed in section V. In the absence of antiferromagnetic and ferromagnetic exchange fields, the lattice structure has nonsymmorphic symmetry ${C_{2x}|\frac{1}{2}0}$, where $C_{2x}$ is the two-fold screw symmetry and $(\frac{1}{2}0)$ is half of the lattice translation in $\hat{x}$ direction. The presence of the antiferromagnetic exchange field breaks the time-reversal symmetry $\mathcal{T}$, but the combination of $\mathcal{T}$ and the nonsymmorphic symmetry is preserved. In this case, the pristine AF XMnY QL (i.e., $\lambda_{FM}=0$) is a topological insulator, as the parameters satisfy the relation $-4t_{I}<\lambda_{AF}<0$. As $\lambda_{AF}$ becomes positive or smaller than $-4t_{I}$, a phase transition to a trivial insulator occurs. In the presence of a ferromagnetic substrate (i.e., $\lambda_{FM}\ne0$), the combination of $\mathcal{T}$ and the nonsymmorphic symmetry is broken so that the TI phase is broken. The helical edge states in the nanoribbons were gapped.

The proximity to the s-wave superconductor induces superconductivity in AF XMnY QL. According to the effective BdG theory, the Hamiltonian in the Nambu space is:
\begin{equation}
H_{BdG}=\begin{bmatrix}
H_{QL}(\mathbf{k})-\mu & H_{\Delta} \\
H_{\Delta}^{\dag} & -H_{QL}^{T}(-\mathbf{k})+\mu
\end{bmatrix}\label{eq_BdGhamil}
\end{equation}
where $\mu$ is the chemical potential and $H_{\Delta}=-i\sigma_{y}\tau_{A}\Delta_{A}-i\sigma_{y}\tau_{B}\Delta_{B}$ is the superconducting pairing with $\tau_{A(B)}=(\tau_{0}\pm\tau_{z})/2$ and $\Delta_{A(B)}$ being the superconducting pairing parameter at the A(B) sublattice. The quasi-particle spectrum was obtained by diagonalization of the Hamiltonian in the first Brillouin zone. The topological properties of the bulk band are described by the BdG Chern number, which is given by the integration of the Berry curvature in the first Brillouin zone. The Berry curvature of the mth band at wave vector $\mathbf{k}$ is given as follows:
\begin{eqnarray}
&&\Omega^{m}_{BdG}(\mathbf{k})=\label{eq_berrycurvature}\\
&&\sum_{n\ne m}\frac{-2Im\langle\Psi_{\mathbf{k}}^{m}|\partial_{k_{x}}H_{BdG}(\mathbf{k})|\Psi_{\mathbf{k}}^{n}\rangle\langle\Psi_{\mathbf{k}}^{n}|\partial_{k_{y}}H_{BdG}(\mathbf{k})|\Psi_{\mathbf{k}}^{m}\rangle}{(E_{\mathbf{k}}^{m}-E_{\mathbf{k}}^{n})^{2}}\nonumber
\end{eqnarray}
where $E_{\mathbf{k}}^{m}$ and $|\Psi_{\mathbf{k}}^{m}\rangle$ are the eigenvalues and eigenstates of the mth band at wave vector $\mathbf{k}$. Thus, the BdG Chern number is given as:
\begin{equation}
C_{BdG}=\frac{1}{2\pi}\sum_{m}^{occ}\int_{BZ}d^{2}\mathbf{k}\Omega^{m}_{BdG}(\mathbf{k})
\end{equation}
where summation covers the occupied band \cite{Petra2020}. The number of pairs of chiral Majorana edge states in the nanoribbons is equal to $C_{BdG}$. In the absence of the superconductor substrate (i.e., $\Delta_{A}=\Delta_{B}=0$), the Hamiltonian is decomposed into sub-matrix for electron and hole with opposite Chern number, so that $C_{BdG}$ is always zero. As $\Delta_{A(B)}$ become nonzero, the coupling between electron and hole could induce topological phase transition to the phase with $C_{BdG}$ being nonzero.

Two types of nanoribbons, designated as parallel and diagonal nanoribbons, were considered, and their lattice structures are plotted in Fig. \ref{figure_system}(b) and (c). For a parallel nanoribbon, the periodic direction is along the $\Gamma-X$ direction of the bulk ($\hat{x}$ in Fig. \ref{figure_system}(b)), and the width direction is along the $\hat{y}$ direction. For a diagonal nanoribbon, the periodic direction is along the $\Gamma-M$ direction of the bulk ($\hat{x}+\hat{y}$ in Fig. \ref{figure_system}(c)), and the width direction is along the $\hat{x}-\hat{y}$ direction. The superunit cell along the width direction was selected to construct the Hamiltonian of the tight-binding model. For the pristine AF XMnY QL in the topological insulating phase, the parallel nanoribbons host two pairs of helical edge states, but the diagonal nanoribbon does not host a gapless edge state owing to the breaking of nonsymmorphic symmetry \cite{Chengwang2020}. As the superconductivity pairing parameters increased, the band structures of the particles and holes coupled with each other. As the bulk is the topological phase with nonzero $C_{BdG}$, chiral Majonara edge states appear. There are $C_{BdG}$ forward (backward) moving edge states localized at the bottom (top) edge of the nanoribbon, as indicated by the bule (red) arrows in Fig. \ref{figure_system}(b) and (c). Because the topological superconductor phase is not protected by nonsymmorphic symmetry, it is protected by the bulk band topology with a nonzero BdG Chern number, and the chiral Majonara edge states appear in both parallel and diagonal nanoribbons. The presence of the chiral Majonara edge states require the broken of either time-reversal symmetry or pseudospin symmetry in the Nambu space, so that either $\lambda_{FM}\ne0$ or $\Delta_{A}\ne\Delta_{B}$ is required for the topological superconducting phase.

\section{Phase diagram with varying superconducting pairing parameters}

\begin{figure}[tbp]
\scalebox{0.58}{\includegraphics{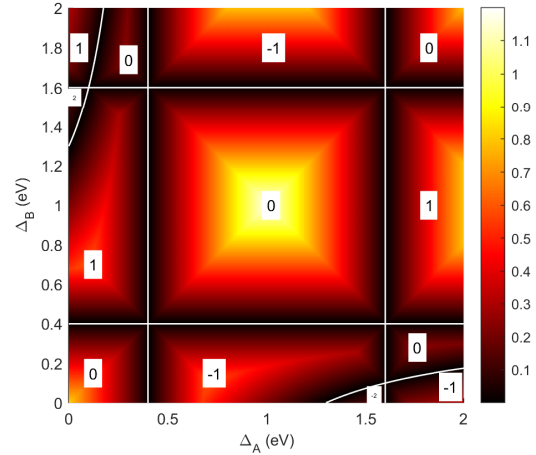}}
\caption{ Band gap and BdG Chern number of the AF XMnY QL in proximity with the s-wave superconductor with varying pairing parameters $\Delta_{A}$ and $\Delta_{B}$, without the ferromagnetic substrate (i.e. $\lambda_{FM}=0$). The AF parameter is $\lambda_{AF}=0.4$ eV. The color scale represents the bulk band gap. The boundaries where the bulk band gap closes at the high symmetric points are plotted as white (solid) lines. }
\label{figure_phase0}
\end{figure}

In this section, we consider the AF XMnY QL in proximity to the s-wave superconductor but in the absence of a ferromagnetic substrate. We first demonstrate the system with $\lambda_{AF}=0.4$ eV, $\lambda_{FM}=0$ eV and $\mu=0$ eV. For varying $\Delta_{A}$ and $\Delta_{B}$, the bulk bandgap is plotted in Fig. \ref{figure_phase0}. As $\Delta_{A}$ and $\Delta_{B}$ vary, the bulk band gap can approach one of the highly symmetric points in the first Brillouin zone, which triggers a topological phase transition. The condition of the phase boundary can be obtained by diagonalizing the BdG Hamiltonian (\ref{eq_BdGhamil}) at one of the highly symmetric points. At $X$ or $M$, with $\mathbf{k}=(\pi,0)$ or $\mathbf{k}=(\pi,\pi)$, the coefficients of the first two terms in the Hamiltonian are $\tilde{M}=0$, the eight eigenvalues of the BdG Hamiltonian are $\pm\Delta_{A(B)}\pm\lambda_{AF}$ and $\pm\Delta_{A(B)}\pm(4t_{I}+\lambda_{AF})$, respectively. Thus, the band gap closes at the $X$ or $M$ point when the pairing parameters satisfy the relation $\Delta_{A(B)}=\lambda_{AF}$ or $\Delta_{A(B)}=4t_{I}+\lambda_{AF}$, respectively. At $\Gamma$ or $Y$ point with $\mathbf{k}=(0,0)$ or $\mathbf{k}=(0,\pi)$, the coefficients of the first two terms in the Hamiltonian is $\tilde{M}=2(t_{1}+t_{2})$ or $\tilde{M}=t_{1}-t_{2}$, so that the eight eigenvalue of the BdG Hamiltonian is $\pm\frac{1}{2}(\Delta_{A}-\Delta_{B})\pm\frac{1}{2}\sqrt{[\Delta_{A}+\Delta_{B}\pm2(\lambda_{AF}-4t_{I})]^{2}+16(t_{1}+t_{2})^{2}}$ and  $\pm\frac{1}{2}(\Delta_{A}-\Delta_{B})\pm\frac{1}{2}\sqrt{(\Delta_{A}+\Delta_{B}\pm2\lambda_{AF})^{2}+16(t_{1}-t_{2})^{2}}$, respectively. The gap-closing condition is given by the relation in which two of the eigenvalues are zero, that is, $\Delta_{A}\Delta_{B}\pm(\lambda_{AF}-4t_{I})(\Delta_{A}+\Delta_{B})+(\lambda_{AF}-4t_{I})^{2}+4(t_{1}+t_{2})^2=0$ or $\Delta_{A}\Delta_{B}\pm\lambda_{AF}(\Delta_{A}+\Delta_{B})+\lambda_{AF}^{2}+4(t_{1}-t_{2})^2=0$, for the gap closing at $\Gamma$ or $Y$ points, respectively. The phase boundaries are shown in Fig. \ref{figure_phase0} as white (solid) line, which are confirmed by the numerical result of the bulk band gap. If the antiferromagnetic exchange field is negative, for example, $\lambda_{AF}=-0.4$ eV, the phase diagram is similar to that in Fig. \ref{figure_phase0}.

The BdG Chern numbers in each phase regime are shown in Fig. \ref{figure_phase0}. When $\Delta_{A}=\Delta_{B}$, the BdG Chern number is always zero. Phase regimes with a norzero BdG Chern number require the condition $\Delta_{A}\ne\Delta_{B}$. For realistic materials, the condition $\Delta_{A}\ne\Delta_{B}$ can be satisfied by engineering the interface between AF XMnY QL and the superconductor substrate. In a significant portion of the phase space, the phase regimes have a BdG Chern number $\pm1$. In two-phase regimes with $\Delta_{A(B)}$ close to zero and $\Delta_{B(A)}$ near 1.5 eV, where the area is small, the BdG Chern number is $2(-2)$.

For the systems with $\mu\ne0$, the phase boundaries are moved toward the x and y axis. At $X$ and $M$ points, the gap closing conditions are modified to be $\Delta_{A(B)}=\pm\sqrt{\lambda_{AF}^2-\mu^{2}}$ and $\Delta_{A(B)}=\pm\sqrt{(4t_{I}+\lambda_{AF})^2-\mu^{2}}$, respectively. The gap closing condition at $\Gamma$ and $Y$ points are also modified, but the solution is too lengthly to be exhibited. As a result, a smaller value of $\Delta_{A(B)}$ could induce the topological superconducting phase.

\section{Effect of ferromagnetic substrate}

Although a topological superconductor phase with a nonzero BdG Chern number is found, the sublattice nonuniformity of the superconducting pairing parameters is necessary, which increases the difficulty of experimental implementation of the topological superconductor phase. We found that the addition of a ferromagnetic exchange field shifts the phase boundary, so that the topological superconductor phase appears when the superconducting pairing parameters are uniform, that is, $\Delta_{A}=\Delta_{B}$. In this section, we present the numerical result with $\lambda_{FM}=0.2$ eV to demonstrate the physical effect of the additional proximity to the ferromagnetic substrate.

\begin{figure}[tbp]
\scalebox{0.7}{\includegraphics{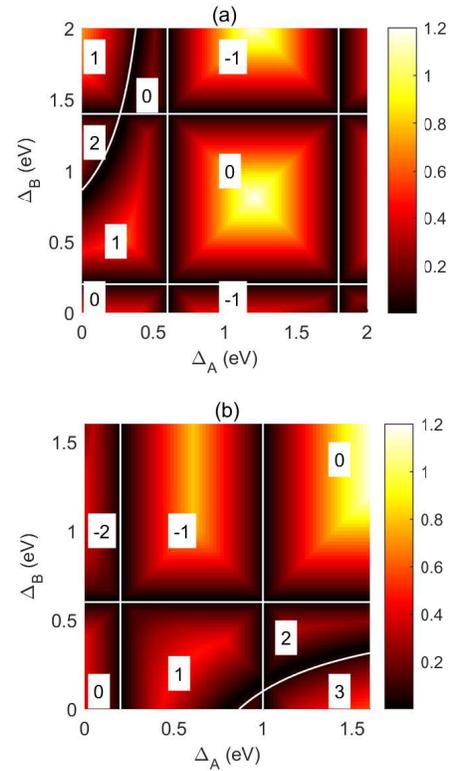}}
\caption{ Band gap and BdG Chern number of the AF XMnY QL in proximity with the s-wave superconductor with varying pairing parameters $\Delta_{A}$ and $\Delta_{B}$, with additional proximity to the ferromagnetic substrate with $\lambda_{FM}=0.2$ eV. The AF parameter is (a) $\lambda_{AF}=0.4$ eV and (b) $\lambda_{AF}=-0.4$ eV. The color scale represents the bulk band gap. The boundaries where the bulk band gap closes at the high symmetric points are plotted as white (solid) lines. }
\label{figure_phase1}
\end{figure}

\begin{figure*}[tbp]
\scalebox{0.55}{\includegraphics{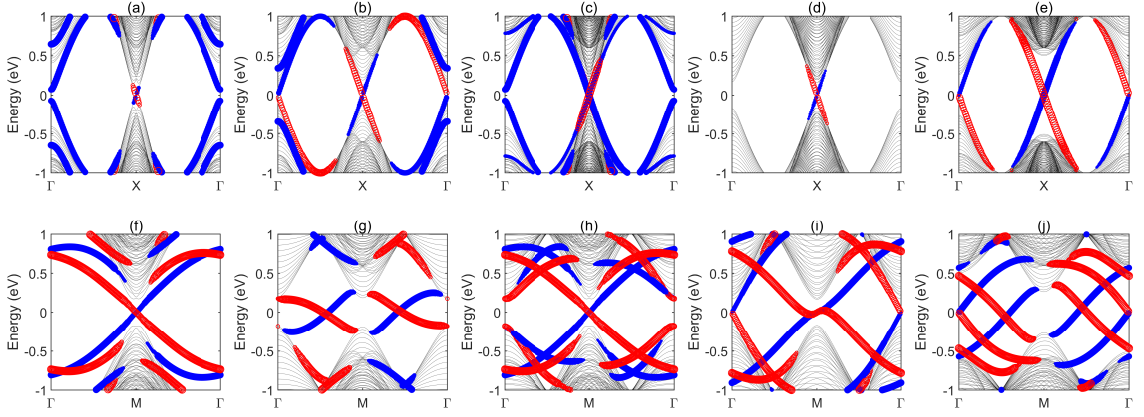}}
\caption{ The band structure of the parallel and diagonal nanoribbon for the top and bottom row, respectively. The size of the blue (filled) and red (empty) dots represent the degree of localization at the bottom and top edges in Fig. \ref{figure_system}(b) and (c). The number of unit cell along the width direction is 50 for both nanoribbons. The system parameters are $\lambda_{FM}=0.2$ eV for all cases, $\lambda_{AF}=0.4$ eV in (a,b,f,g), $\lambda_{AF}=-0.4$ eV in (c,d,e,h,i,j). The superconducting pairing parameters and the corresponding BdG Chern number are (a,f) $\Delta_{A}=\Delta_{B}=0.4$ eV, $C_{BdG}=1$; (b,g) $\Delta_{A}=0$, $\Delta_{B}=1.22$ eV, $C_{BdG}=2$; (c,h) $\Delta_{A}=\Delta_{B}=0.4$ eV, $C_{BdG}=1$; (d,i) $\Delta_{A}=1.2$ eV, $\Delta_{B}=0.45$ eV, $C_{BdG}=2$; and (e,j) $\Delta_{A}=1.5$ eV, $\Delta_{B}=0$, $C_{BdG}=3$. }
\label{figure_ribbon2}
\end{figure*}

We first investigate the system with $\mu=0$ eV. At highly symmetric points, the coefficient of the spin-mixing term in the Hamiltonian (i.e., the term with $t_{R}$) is zero; thus, the BdG Hamiltonian can be decoupled into two submatrices. One of the two submatrix operates on two spin-up electron orbits and two spin-down hole orbits. In the presence of $\lambda_{FM}$, the diagonal terms of the four orbits are added by the same number $\lambda_{FM}$, the four corresponding eigenvalues are shifted by $\lambda_{FM}$. By contrast, the eigenvalues of the other submatrices are shifted by $-\lambda_{FM}$. In particular, the eight eigenvalues at point $X$ are $\lambda_{FM}+\lambda_{AF}\pm\Delta_{A}$, $\lambda_{FM}-\lambda_{AF}\pm\Delta_{B}$, $-\lambda_{FM}-\lambda_{AF}\pm\Delta_{A}$, and $-\lambda_{FM}+\lambda_{AF}\pm\Delta_{B}$, so that the gap-closing condition is $\Delta_{A}=\pm(\lambda_{FM}+\lambda_{AF})$ or $\Delta_{B}=\pm(\lambda_{FM}-\lambda_{AF})$. Similarly, the gap-closing condition at point $M$ is $\Delta_{A}=\pm(\lambda_{FM}+\lambda_{AF}+4t_{I})$ or $\Delta_{B}=\pm(\lambda_{FM}-\lambda_{AF}-4t_{I})$. At the $\Gamma$ point, the eight eigenvalues are $\frac{1}{2}\eta(\Delta_{B}-\Delta_{A})\pm\frac{1}{2}\sqrt{[\Delta_{A}+\Delta_{B}+\zeta\eta2(\lambda_{AF}-4t_{I})]^{2}+16(t_{1}+t_{2})^{2}}-\zeta\lambda_{FM}$ with $\zeta=\pm1$ and $\eta=\pm1$. Thus, the gap-closing condition at point $\Gamma$ is $\Delta_{A}\Delta_{B}+\eta(\lambda_{AF}-4t_{I})(\Delta_{A}+\Delta_{B})+\eta\lambda_{FM}(\Delta_{B}-\Delta_{A})+(\lambda_{AF}-4t_{I})^{2}+4(t_{1}+t_{2})^{2}=0$ with $\eta=\pm1$. Similarly, the gap closing condition at point $Y$ is $\Delta_{A}\Delta_{B}+\eta\lambda_{AF}(\Delta_{A}+\Delta_{B})+\eta\lambda_{FM}(\Delta_{B}-\Delta_{A})+\lambda_{AF}^{2}+4(t_{1}-t_{2})^{2}=0$.
The phase boundaries are shown in Fig. \ref{figure_phase1}, which is confirmed by the numerical result of the bulk band gap.

For $\lambda_{AF}=0.4$, the bulk bandgap and phase boundaries are plotted in Fig. \ref{figure_phase1}(a). Along the line with $\Delta_{A}=\Delta_{B}$, two-phase regimes with $C_{BdG}=\pm1$ appear. Two specific cases, with $C_{BdG}=1$ and $2$, were studied.

(i) When $\Delta_{A}=\Delta_{B}=0.4$ eV, the band structures of the parallel and diagonal nanoribbons are plotted in Fig. \ref{figure_ribbon2}(a) and (f), respectively. Because the BdG Chern number is one, only one pair of gapless chiral Majorana edge states exists in the bulk gap. In the parallel nanoribbon, the forward- and backward-moving chiral Majonara edge states cross at the $X$ point and zero energy. Another pair of edge states localized at the bottom edge has trivial gaps at the $\Gamma$ point. In a diagonal nanoribbon, the gapless chiral Majonara edge states connect the conduction and valence bands with an extra round trip around the first Brillouin zone, as shown in Fig. \ref{figure_ribbon2}(f). A similar feature was observed for the chiral edge states in the diagonal nanoribbon of the optically induced QAH phase of AF XMnY QL with a Chern number of one \cite{maluo2020}.

(ii) As the superconducting pairing parameters become $\Delta_{A}=0$ and $\Delta_{B}=1.22$ eV, where $C_{BdG}=2$, the band structures of the parallel and diagonal nanoribbons are plotted in Fig. \ref{figure_ribbon2}(b) and (g), respectively. The bulk band had two band valleys at $X$ and $\Gamma$ points. In the parallel nanoribbon, one pair of gapless chiral Majonara edge states connects the conduction and valence bands of the $X$ band valley and crosses at the $X$ point and zero energy. Similarly, the other pairs of gapless chiral Majorara edge states were restricted to the $\Gamma$ band valley. In contrast, for the gapless chiral Majorara edge states in the diagonal nanoribbon, each edge band connects the conduction and valence bands in different band valleys.

In the other case, with $\lambda_{AF}=-0.4$, the pristine AF XMnY QL is in the topological insulator phase with two pairs of helical edge states. The bulk bandgap and phase boundaries in the presence of superconductors and ferromagnetic proximity are plotted in Fig. \ref{figure_phase1}(b). Along the line with $\Delta_{A}=\Delta_{B}$, two-phase regimes with $C_{BdG}=\pm1$ appear in the parameter range $-(\lambda_{AF}+\lambda_{FM})\Delta_{A}<\lambda_{AF}+\lambda_{FM}+4t_{I}$. In the parameter range $\Delta_{A}>\lambda_{AF}+\lambda_{FM}+4t_{I}$, two-phase regimes with $C_{BdG}=2$ and $C_{BdG}=3$ appear, which require condition $\Delta_{A}\ne\Delta_{B}$. Three specific cases with $C_{BdG}=1$, $2$ and $3$ were studied.

(i) For $\Delta_{A}=\Delta_{B}=0.4$ eV, the band structures of the parallel and diagonal nanoribbons are plotted in Fig. \ref{figure_ribbon2}(c) and (h), respectively. In a parallel nanoribbon, three pairs of nearly degenerated bands cross at the $X$ point and zero energy. One pair of edge states is the chiral Majonara edge state, whose bands cross at the $X$ point with a zero gap. The forward and backward moving edge states are localized at the bottom and top edges, respectively. The other two pairs of edge states exhibit similar behavior to the helical edge states: in each edge, the forward and backward moving edge states have orthogonal spin expectations. The bands cross at the $X$ point with a small band gap owing to the finite-size effect. In contrast, in the diagonal nanoribbon, only the gapless bands of the chiral Majonara edge state appear at the $X$ point.

(ii) As the superconducting pairing parameters are adjusted to $\Delta_{A}=1.2$ eV and $\Delta_{B}=0.45$ eV, which corresponds to the phase regime with $C_{BdG}=2$, the band structures of the parallel and diagonal nanoribbons are plotted in Fig. \ref{figure_ribbon2}(d) and (i), respectively. In parallel nanoribbons, the two pairs of chiral Majonara edge states have degenerated energy bands, which cross at the $X$ point and zero energy with linear dispersion. In contrast, in the diagonal nanoribbon, one pair of chiral Majonara edge states crosses at the $\Gamma$ point and zero energy with linear dispersion; the other pairs of chiral Majonara edge states cross at the $M$ point with a peculiar band structure. Near the $M$ point, the dispersion of the edge band localized at the top edge has a wavy shape that crosses zero energy three times. Within a small momentum range near the $M$ point, the dispersion of the two edge bands was nearly degenerated. Consequently, within a small energy range near zero energy, the conductance is four times the quantized conductance instead of two, although the BdG Chern number is two. The peculiar band structure is due to the mixing between the edge bands and the bulk bands at the $M$ point, which trigger the topological phase transition. As $\Delta_{B}$ further decreases, the dispersion of the edge band localized at the bottom edge also has a wavy shape; and the bulk band gap at the $M$ point decreases. When $\Delta_{B}$ crosses a critical value that the bulk band gap at the $M$ point closes and reopens, the edge band localized at each edge split into two edge bands that merge into the bulk band at the $M$ point. In this case, the band structure is similar to that in Fig. \ref{figure_ribbon2}(j). The corresponding BdG Chern number of bulk band increases from two to three.

(iii) As the superconducting pairing parameters are adjusted to $\Delta_{A}=1.5$ eV and $\Delta_{B}=0$ which corresponds to the phase regime with $C_{BdG}=3$, the band structures of the parallel and diagonal nanoribbons are plotted in Fig. \ref{figure_ribbon2}(e) and (j), respectively. In a parallel nanoribbon, two pairs of chiral Majonara edge states have degenerated energy bands that cross at the $X$ point and zero energy; the other pair of chiral Majonara edge states cross at the $\Gamma$ point. All edge bands exhibited a linear dispersion near zero energy. In the diagonal nanoribbon, one pair of chiral Majonara edge states that connect to the bulk conduction and valence bands at $M$ point crosses at $\Gamma$ point. The other two pairs of chiral Majonara edge states exhibit behaviors similar to those shown in Fig. \ref{figure_ribbon2}(g).

If the chemical potential is nonzero, the gap closing conditions at $X$ point are modified to be $\Delta_{A}=\pm\sqrt{(\lambda_{FM}+\lambda_{AF})^2-\mu^{2}}$ or $\Delta_{B}=\pm\sqrt{(\lambda_{FM}-\lambda_{AF})^2-\mu^{2}}$, respectively. For the particular cases with $\mu=\lambda_{FM}\pm\lambda_{AF}$, the gap closing condition is $\Delta_{A}=0$  or $\Delta_{B}=0$, so that arbitrarily small value of  $\Delta_{A(B)}$ can induce the topological superconducting phase. For the systems with $\Delta_{A}=\Delta_{B}$ and $0<\Delta_{A(B)}<\sqrt{(\lambda_{FM}-\lambda_{AF}+4t_{I})^2-\mu^{2}}$, the BdG Chern number is one, and the band structures of the nanoribbons are similar to those in Fig. \ref{figure_ribbon2}(a,f).

\section{Parameters for Realistic Materials}

In the two previous sections, the phase diagrams of the theoretical model with specific parameters are given. Because the bulk band gap is large, the finite size effect in the nanoribbon with small width is negligible. If the heterostructures are consisted of realistic materials, the parameters need to be modified. Previous theoretical calculation predicted that the band gap of varying types of AF XMnY QL ranges from 66 meV to 186 meV. Thus, the parameter $\lambda_{AF}$ can be engineered around 100 meV by choosing varying types of AF XMnY QL.

The FM substrate of the heterostructure in Fig. \ref{figure_system} could be chosen from varying type of ferromagnetic insulator, such as EuO\cite{fmsubstrate01}, EuS\cite{fmsubstrate02}, and BiFeO$_{3}$\cite{fmsubstrate03}. Because the AF XMnY QL have square lattice structure, proximity with ferromagnetic substrate with square lattice structure could be more stable. In Ref \cite{NingMao20}, the band structures and exchange energy of varying types of ABC compounds with square lattice structure with A = Li, Na, K, Rb, Cs, Mg, Ca, Sr, Y, La, Ce, Pr, Nd, Tb, Dy, Ho, Yb, Hf, V, Fe, Co, Cu, Al, Ga, B = V, Mn, Fe, Co, and C = Si, P, Ge, As, Sn, Sb, Pb, Bi, are calculated. Some of the ABC compounds are found to have ferromagnetic order instead of antiferromagnetic order. The band gap of the FM ABC compounds are also as large as 100 meV. Proximity between FM ABC compounds and AF XMnY QL could induce FM exchange field in the AF XMnY QL. As a result, by choosing varying type of ferromagnetic ABC compounds as substrate, the parameters $\lambda_{FM}$ can be engineered.

The superconductor substrate could be chosen from conventional superconductors with square lattice structure. Because the critical temperature of normal superconductors is lower than 39 K, the superconducting gap is less than 3.4 meV. If the parameter $|\lambda_{FM}\pm\lambda_{AF}|$ is not less than 3.4 meV after the materials engineering, the chemical potential need to be engineered within the range of $(0,100)$ meV, so that $\sqrt{(\lambda_{FM}\pm\lambda_{AF})^2-\mu^{2}}$ is smaller than 3.4 meV.

\section{Conclusion}

In conclusion, the heterostructure consisting of a ferromagnetic subatrate/AF XMnY QL/superconductor could be in the topological superconductor phase with varying BdG Chern numbers. The ferromagnetic substrate raises the feasibility of the theoretical model in realistic materials because the condition $\Delta_{A}\ne\Delta_{B}$ is not required. In a large phase regime, the BdG Chern number is equal to one, so only one pair of chiral Majonara edge states appears in the nanoribbons. In a certain phase regime, chiral Majonara edge states coexist with helical edge states. Systems with one pair of chiral Majonara edge states could be robust candidates for implementing non-Abelian braiding statistics.

\begin{acknowledgments}
This project was supported by the Natural Science Foundation of Guangdong Province, China (Grant no. 2022A1515011578), the Project of Educational Commission of Guangdong Province of China (Grant No. 2021KTSCX064), the startup grant at Guangdong Polytechnic Normal University (Grant No. 2021SDKYA117), and the National Natural Science Foundation of China (Grant No. 11704419).
\end{acknowledgments}

\clearpage

\end{document}